\documentclass{emulateapj}
\usepackage{graphicx}
\usepackage{xspace}

\newcommand{\bb}[1]{\ifmmode \mbox{\boldmath $ #1$} \else  \mbox{\boldmath $#1$} \fi}

\newcommand{\dd}{\ensuremath{\,\mathrm{d}}}

\newcommand{\U}[1]{\ensuremath{\mathrm{~#1}}}

\newcommand{\Myr}{\U{Myr}}

\newcommand{\pc}{\U{pc}}
\newcommand{\kpc}{\U{kpc}}

\newcommand{\msun}{\U{M}_{\odot}}
\newcommand{\Msun}{\msun}
\newcommand{\cc}{\U{cm^{-3}}}
\newcommand{\kms}{\U{km\ s^{-1}}}

\newcommand{\hi}{H{\sc i} }

\newcommand{\htwo}{H\ensuremath{_2}}

\newcommand{\tff}{\ensuremath{t_\mathrm{ff}}}
\newcommand{\rhol}{\ensuremath{\rho}}
\newcommand{\rhon}{\ensuremath{\bar{\rho}}}
\newcommand{\rhot}{\ensuremath{\rho_0}}
\newcommand{\rhosfr}{\ensuremath{\rho_\mathrm{SFR}}}
\newcommand{\sigmag}{\ensuremath{\Sigma}}
\newcommand{\sigmasfr}{\ensuremath{\Sigma_\mathrm{SFR}}}

\newcommand{\pdf}{\ensuremath{f_\mathrm{\sigma}}}
\newcommand{\mach}{\ensuremath{\mathcal{M}}}

\newcommand{\tffsat}{\ensuremath{t_\mathrm{s}}}
\newcommand{\esat}{\ensuremath{\epsilon_\mathrm{s}}}

\everymath={\displaystyle}

\newcommand{\eqn}[2][]{Equation#1~\ref{eqn:#2}} %% for plural form, use: \eqn[s]{emc2} and (\ref{eqn:emc3})     to get  Equations (1) and (2)
\newcommand{\fig}[2][]{Figure#1~\ref{fig:#2}}

\renewcommand{\eqn}[2][]{equation#1~(\ref{eqn:#2})}
\renewcommand{\fig}[2][]{Fig#1.~\ref{fig:#2}}

\newcommand{\new}[1]{#1}
\begin{document}

\title{Star formation laws and thresholds from ISM structure and turbulence}
\shorttitle{Star formation laws and thresholds from ISM structure and turbulence}

\author{Florent~Renaud, Katarina Kraljic and Fr\'ed\'eric~Bournaud}
\shortauthors{Renaud, Kraljic \& Bournaud}
\affil{Laboratoire AIM Paris-Saclay, CEA/IRFU/SAp, Universit\'e Paris Diderot, F-91191 Gif-sur-Yvette Cedex, France}
\email{florent.renaud@cea.fr}

%%%%%%%%%%%%%%%%%%%%%%%%%%%%%%%%%%%%%%%%%%%%%%%%%%
\begin{abstract}
We present an analytical model of the relation between the surface density of gas and star formation rate in galaxies and clouds, as a function of the presence of supersonic turbulence and the associated structure of the interstellar medium. The model predicts a power-law relation of index 3/2, flattened under the effects of stellar feedback at high densities or in very turbulent media, and a break at low surface densities when ISM turbulence becomes too weak to induce strong compression. This model explains the diversity of star formation laws and thresholds observed in nearby spirals and their resolved regions, the Small Magellanic Cloud, high-redshift disks and starbursting mergers, as well as Galactic molecular clouds. While other models have proposed interstellar dust content and molecule formation to be key ingredients to the observed variations of the star formation efficiency, we demonstrate instead that these variations can be explained by interstellar medium turbulence and structure in various types of galaxies.
\end{abstract}
\keywords{stars: formation --- galaxies: ISM --- hydrodynamics --- methods: analytical}

%%%%%%%%%%%%%%%%%%%%%%%%%%%%%%%%%%%%%%%%%%%%%%%%%%
\section{Introduction}

Galactic-scale star formation (SF) laws are not universal. Galaxy mergers, when they experience a starburst phase, convert their gas into stars within a depletion time up to ten times shorter than spiral galaxies \citep{Daddi2010b, Genzel2010, Saintonge2012}. Oppositely, SF in dwarf galaxies is less efficient than in spirals \citep{Leroy2008, Bolatto2011}. On smaller scales, molecular clouds may follow similar scaling laws but with shorter conversion timescales \citep[e.g.][]{Lada2010}.

To explain these environment-dependent differences, some models have emphasized the role of interstellar medium (ISM) chemistry (metallicity and dust content) in triggering gas cooling and molecule formation. The transition from long to short gas depletion times may correspond to the critical column-density needed to shield gas from the ambient ultraviolet radiation \citep{Schaye2004} and/or to efficiently convert \hi into \htwo\ \citep{Krumholz2009a}, which strongly depends on the metallicity. This is supported by recent observations of the Small Magellanic Cloud (SMC), of lower metallicity than spirals, where the star formation rate (SFR) surface density hardly reaches the standard regime even at relatively high gas surface densities \citep{Bolatto2011}. However, such descriptions do not explain the very long gas depletion timescales in metal-rich and molecule-rich ellipticals \citep{Saintonge2012}. Furthermore, significant SF activity can be found in regions where the molecular fraction is low \citep{Boissier2008}. More generally, variations in the SF scaling-laws about as strong in \htwo\ as in total gas have been detected, suggesting that molecule formation alone does not drive these variations \citep{Saintonge2012}.

Nevertheless, other factors may explain the observed scaling laws. The gas surface density observed in an entire galaxy or a relatively large region is a global property that results from a very heterogeneous distribution of the volume density on small scales. The latter varies broadly from quasi-empty holes to dense clouds and cores, under the effect of the ISM turbulence, which is supersonic for a large fraction of the mass \citep[e.g.][]{Audit2010}. \citet{Elmegreen2002b} has shown that the turbulent structure of the ISM can naturally explain the \citet{Kennicutt1998b} relation for spiral galaxies.

In this paper, we develop an analytic model describing the variations and thresholds in SF laws depending on the supersonic nature of the turbulence and the resulting structure of the ISM in several types of galaxies.

%%%%%%%%%%%%%%%%%%%%%%%%%%%%%%%%%%%%%%%%%%%%%%%%%%%%%%%%%%%%%%%%%%%%%%%%%%%%%%%%
\section{Theory}

%%%%%%%%%%%%%%%%%%%%%%%%%%%%%
\subsection{Analytical formalism}
We consider a region of surface $S$ and thickness $h$, whose total mass $M$ is distributed according to a mass-weighted probability density function (PDF) $f$. The volume of the region can be written as the sum of the volumes occupied by the gas at all possible densities $\rhol$, i.e.
\begin{equation}
hS = \int_0^{\infty} \frac{M f(x)}{\rhol} \dd x = \frac{M}{\rhon} \int_0^{\infty} \frac{f(x)}{x} \dd x,
\end{equation}
where $x = \rhol / \rhon$ is a normalization of the \emph{local} gas volume density $\rhol$ (i.e. for scales $\leq 1 \pc$), related to gas surface density of the region $\sigmag = M/S$ through
\begin{equation}
\label{eqn:rhobar}
\rhon = \frac{\sigmag}{h} \int_0^{\infty} \frac{f(x)}{x} \dd x.
\end{equation}

The SFR surface density reads
\begin{equation}
\sigmasfr = \frac{1}{S} \int_{0}^{\infty} \frac{M f(x)}{\rhol} \rhosfr \dd x,
\end{equation}
where $\rhosfr$ is the local SFR volume density. Therefore, the general expression of the SFR surface density is\footnote{See also \citealt{Padoan2011} and \citealt{Hennebelle2011} for comparable approaches.}
\begin{equation}
\label{eqn:sigmasfr}
\sigmasfr = h \frac{\int_{0}^{\infty} f(x) x^{-1} \rhosfr \dd x}{\int_0^{\infty} f(x) x^{-1} \dd x}.
\end{equation}

%%%%%%%%%%%%%%
\subsubsection{Log-normal PDF}

Most of the ISM mass is supersonically turbulent, which generates a log-normal PDF \citep[e.g.][]{Vazques1994, Nordlund1999, Wada2001}:
\begin{equation}
\pdf(x) = \frac{1}{x\sqrt{2\sigma^2\pi}}\exp{\left[-\frac{\left(\ln{(x)}-\frac{\sigma^2}{2}\right)^2}{2\sigma^2}\right]},
\end{equation}
whose dimensionless width $\sigma$ is related to the Mach number $\mach$ (and thus to the gas velocity dispersion) through $\sigma^2 \approx \ln{(1+ 3\mach^2/4)}$. This was primarily found for isothermal gas, but models of non-isothermal ISM including gravity and stellar feedback have shown that deviations from the log-normal functional form are negligible, at least in isolated disk galaxies \citep{Tasker2009, Bournaud2011b}.

%%%%%%%%%%%%%%
\subsubsection{Dynamical star formation}
\label{sec:sf}

At small-scale, cold enough gas becomes supersonically turbulent and hosts the shocks which trigger the process of star formation. This typically happens above a certain volume density $\rhot$: in nearby spirals for instance, the velocity dispersion of $6 \-- 10 \kms$ requires cooling below $\sim 10^4 \U{K}$ for ISM turbulence to become supersonic. This only becomes possible above $\rhot \approx 10 \cc$, \new{at solar metallicty, according to calculations at galactic scale \citep{Bournaud2010b}, and detailed ISM models including developed turbulence and radiative transfer which show $\mach \approx (\rho / 10 \cc)^{1/2}$ \citep[Fig.~4 and 9]{Audit2010}}. Above this density threshold, a fraction of the gas becomes gravitationally unstable and collapses, converting a constant fraction\footnote{Observations suggest $\epsilon \approx 0.01$, independently of the local density and scale \citep{Krumholz2007a}. We adopt this value throughout the paper.} $\epsilon = 0.01$ of its mass into stars per free-fall time $\tff$, as supported by observations \citep{Elmegreen2002b}:
\begin{equation}
\label{eqn:sfr}
\rhosfr = \left\{\begin{array}{ll}
0  & \textrm{ if } \rhol \leq \rhot \\
\epsilon \frac{\rhol}{\tff} = \epsilon \sqrt{\frac{32G}{3\pi}} \rhol^{3/2} & \textrm{ else}\\
\end{array}\right..
\end{equation}

%%%%%%%%%%%%%%
\subsubsection{Regulation by stellar feedback}

\new{In dense cores, stellar feedback limits the conversion of the gas into stars, by heating, ionizing and even ejecting the gaseous left-overs. This gas becomes available again for SF after a time $\tffsat$. No more than $\esat = 30\%$ of the gas mass can be consumed for SF per timescale $\tffsat$ \citep{Bontemps1996, Matzner2000}. This translates into a saturation in the local SF law:}
\begin{equation}
\label{eqn:feedback}
\rhosfr = \left\{\begin{array}{ll}
0  & \textrm{ if } \rhol \leq \rhot \\
\min{\left( \epsilon \sqrt{\frac{32G}{3\pi}} \rhol^{3/2} , \esat \frac{\rhol}{\tffsat} \right)} & \textrm{ else}
\end{array}\right..
\end{equation}
\new{The (re)formation of star forming clouds is generally triggered by galactic-scale processes (e.g. \citealt{Dobbs2009} for spirals, \citealt{Teyssier2010} for mergers and \citealt{Bournaud2007} at high redshift), therefore $\tffsat = 100 \Myr$ over a broad range of galaxy masses. For instance, this represents the interval between the compression by two spiral arms, or for giant clumps to collapse in high-redshift disks. In nuclear starbursts however, the dynamical timescale is much shorter ($\sim 10 \Myr$) but strong stellar feedback from OB-type stars takes over and also regulates SF, by limiting the conversion of gas into stars to $\esat \approx 0.3$ \citep{Murray2010} over the duration of the starburst event, i.e. $\tffsat = 100 \Myr$ here again \citep{DiMatteo2008}.}

%%%%%%%%%%%%%%%%%%%%%%%%%%%%%
\subsection{Results}

\begin{figure}
\includegraphics{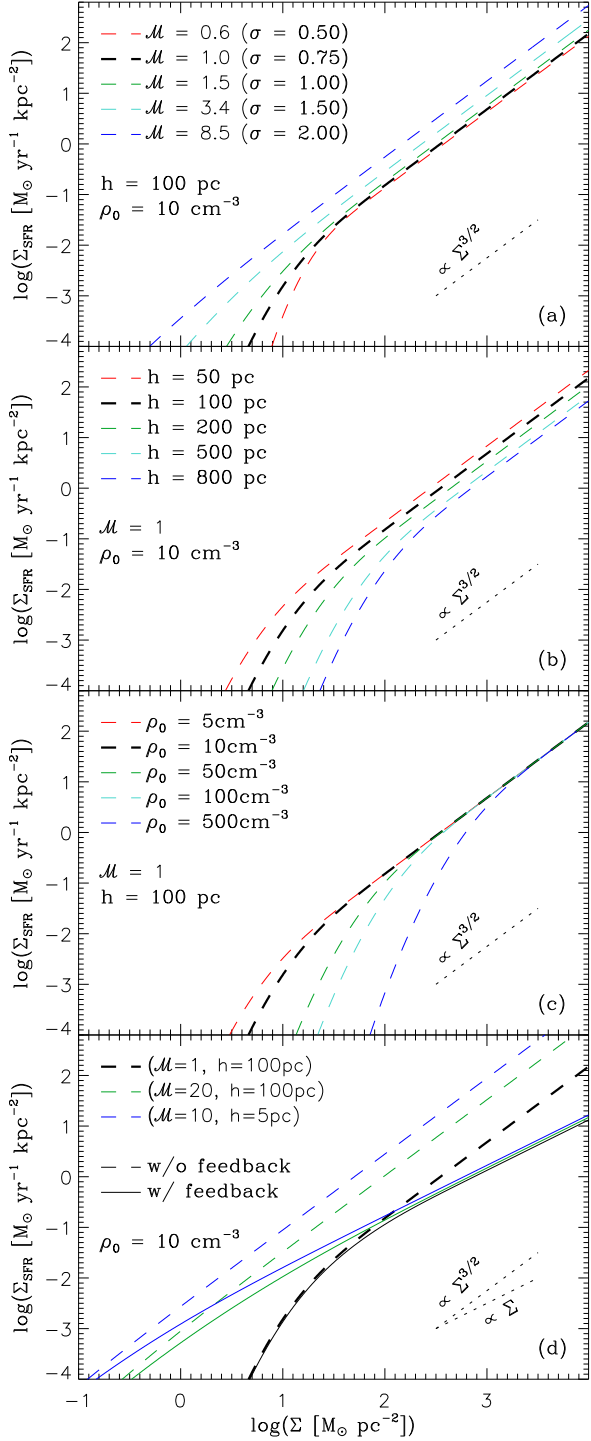}
\caption{Surface density of the star formation rate, computed from \eqn{sigmasfrlognormal}, with $\epsilon = 0.01$. The dependences with the Mach number $ \mach$ (via the width $\sigma$ of the PDF, panel a), the thickness $h$ (panel b) and the density threshold $\rhot$ (panel c) are shown. Panel (d) shows the regulation of star formation due to the stellar feedback (solid lines, equation~\ref{eqn:sfrsat}), compared to no regulation, as above (dashed lines). The black dashed line is the same in all panels.}
\label{fig:sfr}
\end{figure}

\emph{Without feedback} (equation~\ref{eqn:sfr}) and for a log-normal PDF, we obtain\footnote{$\mathrm{erfc} : \xi \mapsto \frac{2}{\sqrt{\pi}} \int_\xi^{\infty} \exp{(-t^2)} \dd t = 1- \textrm{erf}(\xi).$}
\begin{equation}
\label{eqn:sigmasfrlognormal}
\sigmasfr = \epsilon \sqrt{\frac{8G}{3\pi}} \frac{\exp{\left(\frac{3\sigma^2}{8}\right)}}{\sqrt{h}} \sigmag^{3/2}  \mathrm{erfc}\left(\frac{\ln{\left(\frac{\rhot h}{\sigmag}\right)}-\sigma^2}{\sigma\sqrt{2}}\right).
\end{equation}
As illustrated in \fig{sfr}.abc, this corresponds to a $3/2$-index power law which, for any non-zero threshold $\rhot$, falls off at low surface densities. Hence, the combination of standard functional forms for the density PDF of the ISM and the local SFR naturally results in a threshold (or ``break'') in $\sigmasfr$, the shape and position of which depend on $\mach$, $h$ and $\rhot$.

As shown in \fig{sfr}.d, the regulation by feedback (equation~\ref{eqn:feedback}) induces a shallower power law (of index unity) at high surface densities:
\begin{equation}
\label{eqn:sfrsat}
\sigmasfr = b_{\sigma,0}\ \sigmag^{3/2} + c_{\sigma,0}\ \sigmag,
\end{equation}
with the functional forms:
\begin{equation}
\begin{array}{ll}
b_{\sigma,\delta} =& \epsilon \sqrt{\frac{8G}{3\pi}} \frac{\exp{\left(\frac{3\sigma^2}{8}\right)}}{\sqrt{h}} \left[ \mathrm{erfc}\left(\frac{\ln{\left(\frac{\rhot h}{\sigmag}\right)}-\sigma^2 - \delta}{\sigma\sqrt{2}}\right) \right.\\ 
& \left. - \mathrm{erfc}\left(\frac{\ln{\left(\frac{3\pi\esat^2 h}{32 G \tffsat^2 \epsilon^2\sigmag}\right)}-\sigma^2 - \delta}{\sigma\sqrt{2}}\right) \right]\\
c_{\sigma,\delta} =& \frac{\esat}{2\tffsat} \mathrm{erfc}\left(\frac{\ln{\left(\frac{3\pi\esat^2 h}{32 G \tffsat^2 \epsilon^2\sigmag}\right)}-\frac{\sigma^2}{2} - \delta}{\sigma\sqrt{2}}\right).
\end{array}
\end{equation}
The transition from the standard regime to this feedback-regulation one is shifted toward lower $\sigmag$ for high $\mach$ or, in a more modest way, small $h$.

%%%%%%%%%%%%%%%%%%%%%%%%%%%%%%%%%%%%%%%%%%%%%%%%%%%%%%%%%%%%%%%%%%%%%%%%%%%%%%%
\section{Comparison to observed SF laws}

%%%%%%%%%%%
\subsection{Local spirals and the Small Magellanic Cloud}

\begin{figure}
\includegraphics{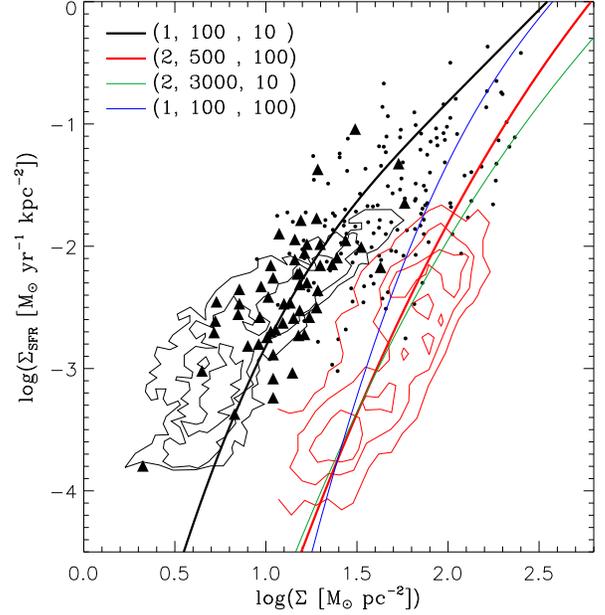}
\caption{Star formation rate surface density computed with parameters ($\mach$, $h$ in$\pc$, $\rhot$ in$\cc$) representative of spiral galaxies (black) and of the Small Magellanic Cloud (SMC, red). The models are compared to observational data of the THINGS survey \citep[black contours,][]{Bigiel2008}, other spiral galaxies \citep[triangles,][]{Kennicutt1998b}, M51 regions \citep[black dots,][]{Kennicutt2007} and the SMC \citep[red contours,][]{Bolatto2011}. The blue and green lines are alternative fits of lower quality (see text).}
\label{fig:sfr_mw_smc}
\end{figure}

In typical local spiral galaxies, large-scale \hi and CO reservoirs have velocity dispersions of $6 \-- 10 \kms$ \citep{Combes2002} which, for temperatures of $10^{3\-- 4} \U{K}$, corresponds to $\mach \approx 1$ and $h \approx 100 \pc$. With these values, our model naturally reproduces the break in SF law observed in entire galaxies and smaller regions (but yet $\ge h$), as illustrated in \fig{sfr_mw_smc}.

The SMC has different properties: we estimate\footnote{The thickness of the SMC is derived using \citet[Chapter 1]{Combes2002} with a velocity dispersion of $\approx 20 \kms$ and a rotation speed of $40 \kms$ at the \hi half-mass radius \citep[$1.5 \kpc$,][]{Stanimirovic2004}.} its thickness to be $\sim 500\pc$, but note that its interaction with the Large Magellanic Cloud could make it even thicker \citep{Besla2012}. Furthermore, the low metallicity of the SMC \citep[$\sim 0.2 Z_\odot$, e.g.][]{Bolatto2008} implies a less efficient cooling of the ISM leading to a transition to supersonic turbulence at $\rhot = 100 \cc$ \citep[using cooling calculations similar to those of][]{Bournaud2010b}. Using these values, our model matches the observations (\fig{sfr_mw_smc}, red).

The (relative) uncertainty on $h$ allows us to adjust the triplet of parameters to explore other possibilities. First, considering that $\rhot$ is independent of the metallicity (and thus has the same value as for spirals) would require adjusting the thickness to an unrealistic value ($3\kpc$) to match the data (green curve in \fig{sfr_mw_smc}). Second, not accounting for the structure of the ISM ($\mach$ and $h$) and only focusing on the role of the onset of supersonic turbulence through a high $\rhot$ would lead to a too steep relation (blue line in \fig{sfr_mw_smc}). Alternatively, if we would assume that the formation of \htwo\ triggers SF instead of the transition to supersonic turbulence, we could model this with an even higher threshold ($\rhot \approx 100 \-- 300 \cc$ for the formation of molecules at subsolar metallicity), and thus get a steeper slope failing to fit the observations. Hence, the observed SF law in the SMC is best explained by the onset of supersonic turbulence and the associated ISM structure\footnote{In these regimes of low $\mach$ and $\sigmasfr$ (spiral disks and SMC), the feedback has a very mild effect, not affecting the relations.}.

%%%%%%%%%%%
\subsection{Application to starbursting mergers}
\label{sec:mergers}

\begin{figure}
\includegraphics{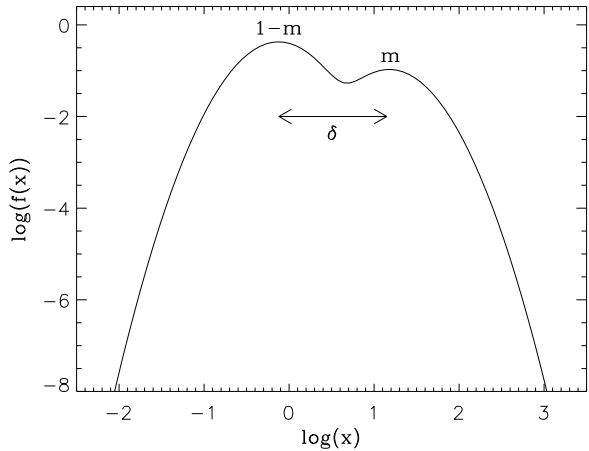}
\caption{Density PDF of a merger with the parameters adopted throughout the paper $\sigma_1 = \sigma_2$, $\delta=3$ and $m=0.2$ (see text).}
\label{fig:twopdf}
\end{figure}

When they experience a starburst phase, galaxy mergers often have a part of their gas compressed in a nuclear disk, which we model by converting a mass fraction $m$ of the initial PDF into the form of a denser component (using a dimensionless parameter $\delta$, see \fig{twopdf}): $(1-m) f_\mathrm{\sigma_1}(x) + m f_\mathrm{\sigma_2}(x / \exp{\delta})$. Hydrodynamic simulations of mergers found such excess of high-density components compared to a log-normal PDF, even outside nuclear disks \citep{Teyssier2010, Bournaud2011a}, and PDFs from these works are well fitted by our arbitrary model using $\delta = 3$ and $m = 0.2$. With feedback, we obtain
\begin{equation}
\label{eqn:sigmasfrmerger}
\begin{array}{ll}
\sigmasfr =& \sigmag^{3/2} \left[ (1-m) b_{\sigma_1,0} + m \exp{\left(\frac{3\delta}{2}\right)} b_{\sigma_2,\delta} \right] \\
& + \sigmag \left[ (1-m) c_{\sigma_1,0} +  m \exp{(\delta)} c_{\sigma_2,\delta} \right].
\end{array}
\end{equation}
Such SF law yields a single break, at lower surface densities than in isolated galaxies, and a flattening at high surface densities due to stellar feedback.

%%%%%%%%%%%
\subsection{Clouds, disks and mergers at low and high redshift}

\begin{figure*}
\includegraphics{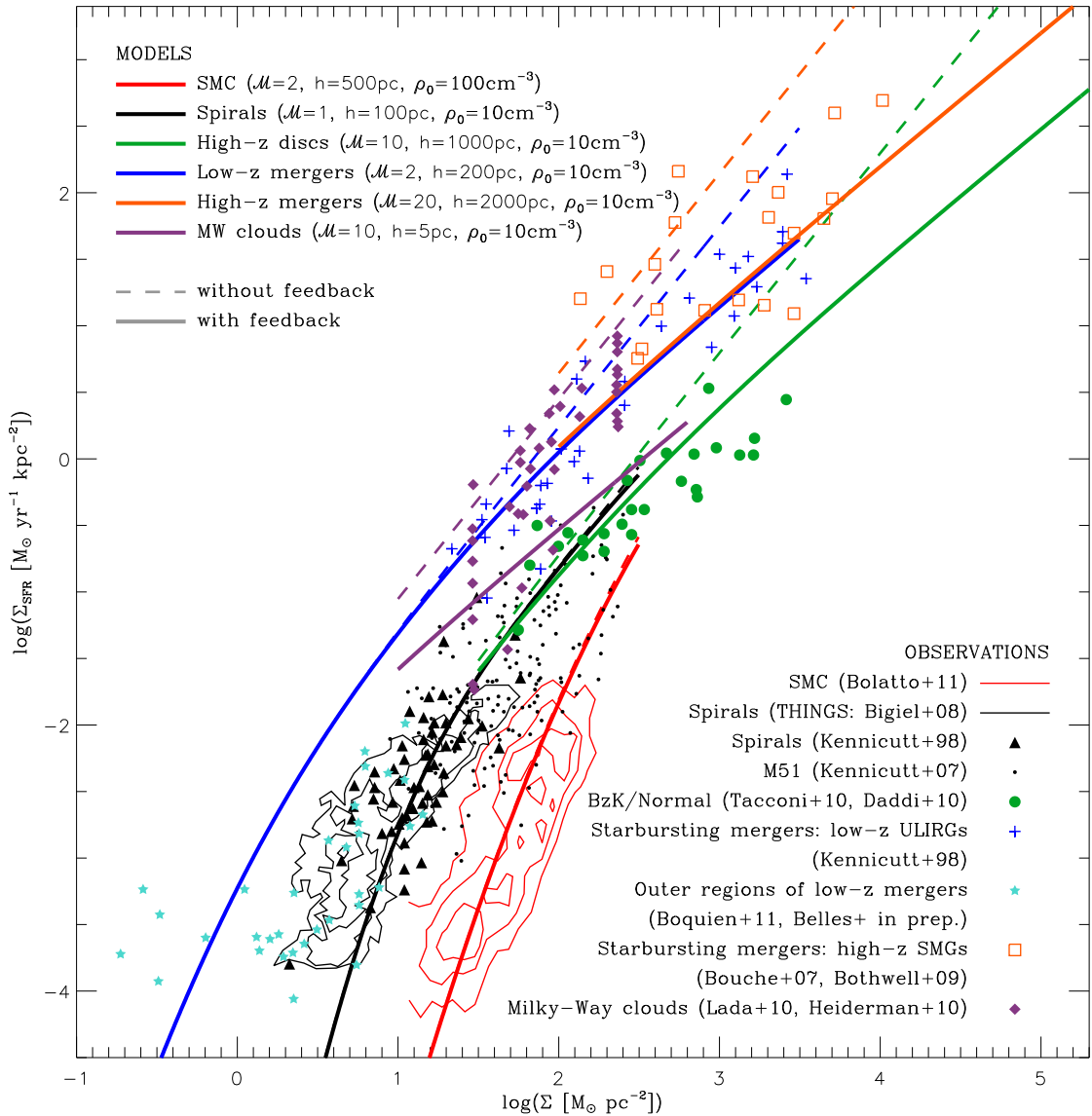}
\caption{Comparison of our models, using physically motivated parameters, with observations of galaxies or regions of galaxies at low and high redshift ($z$).}
\label{fig:sfr_obs}
\end{figure*}
\nocite{Bolatto2011, Bigiel2008, Kennicutt1998b, Kennicutt2007, Tacconi2010, Daddi2010a, Boquien2011, Bouche2007, Bothwell2009, Krumholz2012, Lada2010, Heiderman2010}

High-redshift ($z = 1\-- 2$) disks have strong turbulence \citep[$50 \-- 100 \kms$,][]{Forster2009} leading to a thicker disk \citep[$\sim 1 \kpc$,][]{Elmegreen2006b}, and a higher $\sigmag$ than local spirals. In such a regime, feedback induces a transition of the power-law index from 3/2 to unity, which best matches the data (\fig{sfr_obs}). Note that the onset of supersonic turbulence probably occurs at a lower density threshold than that of nearby spirals ($10 \cc$). However, this would mainly modify the shape of the break, in a range of surface densities much lower than that of the (non-resolved) high-$z$ galaxies considered here.

The ISM in starbursting mergers is at least twice more turbulent than in isolated galaxies (at low-redshift: \citealt{Irwin1994,Elmegreen1995} and high-redshift: \citealt{Bournaud2011b}). This can increase the scale-height of gas reservoirs, although gas in the most central regions is not necessarily thicker than in disk galaxies, because of its high surface density \citep{Downes1998}. In \fig{sfr_obs}, we adopt doubled $\mach$ and $h$ compared to disks at the same redshift. In these dense and turbulent systems, the regulating role of feedback is already significant at $z=0$ and increases with $z$, explaining the deviation from the $3/2$-slope toward the shallower slopes ($1.2 \-- 1.3$), as observed by \citet{Daddi2010b} and \citet{Genzel2010}. Furthermore, the same authors noted that SF relations globally raised by $\sim 0.9$ dex for mergers compared to spirals, which our model retrieves well, thanks to the extended PDF (excess of dense gas and strong turbulence).

The strong turbulence in molecular clouds of the Milky-Way (MW) makes them lie close to the low-$z$ mergers. Although our model may slightly underpredict $\sigmasfr$ for these objects, its qualitative behavior compared to spirals and the general trend due to feedback regulation are in good agreement with the observations. In fact, these molecular clouds are distributed between the model predictions with and without feedback, which could indicate a mix of young clouds not yet affected by feedback, and of old ones undergoing strong regulation.

In summary, the regulation by feedback, the turbulence increasing with the redshift, the extended PDF for starbursting mergers and the higher threshold at low metallicities are necessary aspects to retrieve the diversity of the SF laws. Molecule formation could be concomitant with SF, without being the main physical trigger \citep{Glover2012}.

%%%%%%%%%%%%%%%%%%%%%%%%%%%%%%%%%%%%%%%%%%%%%%%%%%%%%%%%%%%%%%%%%%%%%%%%%%%%%%%
\section{Alternative interpretations, possible tests}

\new{In the proposed framework, a local volume density threshold is required to explain an observable surface density break. We have proposed that the density threshold corresponds to the onset of supersonic turbulence, generating shocks that trigger the gravitational instabilities leading to SF. However, other interpretations are possible, keeping the same formalism.}

\new{First, on top of self-gravity, the galactic environment of a cloud can modify its equilibrium, delaying or accelerating the onset of SF. The tidal field and the shear tend to stabilize clouds against collapse \citep{Elmegreen2006c}. Therefore, the threshold for SF could correspond to the critical density needed to initiate collapse. In pressure equilibrium, $\rhot$ would be the density for which tidal, centrifugal (shear) and Coriolis forces balance self-gravity. Although it depends on the galactocentric radius, this tidal density averaged over the galaxy is $\sim 10 \cc$, with limited variations with galaxy type and redshift. Second, support by magnetic fields might stabilize molecular clouds, leading to higher thresholds and shallower slopes in all regimes.}

\new{Both these alternative interpretations would translate into mild variations of the density threshold with the galactic environment. However, if, as we have proposed here, the local density threshold corresponds to the onset of supersonic turbulence, highly turbulent media (e.g. mergers and high redshift disks) would have very low thresholds leading to breaks in the observed SF relation shifted to low surface brightness, close to the detection limits and thus difficult to detect. Preliminary observations of resolved regions of low-$z$ mergers \citep[Belles et al., in prep,][]{Boquien2011} started to probe this regime (see \fig{sfr_obs}): some regions are spiral-like (as expected since not all mergers are starbursting) but $\sim 20\%$ of them have significant $\sigmasfr$ for very low $\sigmag$, well below the break of spirals. Larger samples are needed to tell apart starbursting regions from spiral-like ones, and to confirme the trend suggested here. This could also be tested with resolved observations of $z=2$ disks, to further probe the physical origin of the local density threshold.}

%%%%%%%%%%%%%%%%%%%%%%%%%%%%%%%%%%%%%%%%%%%%%%%%%%%%%%%%%%%%%%%%%%%%%%%%%%%%%%%
\section{Summary and conclusion}

In this Letter, we present an analytical formalism aiming at describing the relation between the surface density of gas and the surface density of SFR observed in several types of galaxies. The only two ingredients of the model are the gas density PDF (representing the turbulence-driven structure of the ISM), and a local SF law, with a threshold due to the onset of supersonic turbulence, plus regulation by stellar feedback. Our main findings are:
\begin{itemize}
\item When integrated over regions of galaxies or entire galaxies, the threshold in the local SF law translates into a break at low surface densities.
\item Above this break, the local SF law directly imprints the global SFR surface density, leading to a $3/2$-index power-law, followed by a unity slope at high surface density, when feedback takes over turbulence as the main regulation agent.
\item The slow nature of star formation in nearby spirals with Gyr-long gas depletion timescales is mostly explained by turbulent regulation (i.e. divergent flows disrupting gas clouds, \citealt{Elmegreen2002b}). Feedback regulation becomes more prominent in denser and more turbulent systems (e.g. high-$z$ galaxies or Galactic molecular clouds) and produces the observed shallower relation for the entire disk population (average slope 1.1-1.3).
\item The different properties of the SMC can explain the onset of supersonic turbulence at higher surface densities than in spirals, and our model retrieves the observed longer depletion times without necessarily invoking molecule formation or shielding by dust.
\item In starbursting mergers, the strong interaction-induced turbulence results in high-efficiency SF, with a break shifted to low densities but tentatively observed in resolved galaxies.
\end{itemize}
\new{By simply using the typical values of ISM properties from the literature for several galactic environments, our model naturally explains the observations, without fitting them nor fine tuning the parameters.} 

\new{The temperature of the gas shielded from ultraviolet radiations and \emph{in thermal equilibrium} would drop when the density reaches $\approx 0.8 \cc$ \citep{Audit2010}, allowing ISM turbulence to become supersonic at these relatively low volume densities. In spiral galaxies, such low threshold would translate into an apparent break at $< 1 \Msun \pc^{-2}$. Instead, according to models including not only ultraviolet radiative transfer but also turbulent compression, the gas slowly cools below $\sim 10^4 \U{K}$ and enters the supersonic regime at about $10 \cc$ \citep{Audit2010}, which naturally explains the observed break at $\sim 10 \Msun \pc^{-2}$.}

Our results emphasize that the supersonic turbulence plays a key role in triggering and regulating galactic-scale star formation, along with feedback in dense and turbulent media, and that the resulting ISM structure can broadly explain the observed star formation laws and thresholds.

%%%%%%%%%%%%%%%%%%%%%%%%%%%%%%%%%%%%%%%%%%%%%%%%%%
\acknowledgments

We thank Pierre-Emmanuel Belles, Pierre-Alain Duc and Elias Brinks for providing us with their data prior to publication and for comments, Avishai Dekel, Bruce Elmegreen, Mark Krumholz and Am\'elie Saintonge for stimulating discussions, and the anonymous referee for a useful report. We acknowledge support from the EC through grant ERC-StG-257720, and the CosmoComp ITN.

%\bibliographystyle{mn2e}
%\bibliography{biblio}

\end{document}